# Controlled modulation of optical energy in the high order Hermite-Gaussian laser modes

Hemant Kumar Meena, Brijesh Kumar Singh*

*Department of Physics, School of Physical Sciences, Central University of Rajasthan, Rajasthan, 305817, Ajmer, India*

## Abstract

The multiple lobes of high order Hermite-Gaussian (HG) laser modes differ in terms of shape, size, and optical energy distribution. Here, we introduce a generic numerical method that redistributes optical energy among the lobes of high order HG modes such that all the identical low intense lobes become both moderate or high intense lobes and vice-versa, in a controlled manner. Further, the modes which consist of only two types of intensity distribution among its multiple lobes are transformed together into all high intense lobes. Furthermore, in some cases, moderate intense lobes together with high intense lobes become high intense lobes, and moderate intense lobes together with low intense lobes become high intense lobes. Such controlled modulation of optical energy may offer efficient and selective utilization of each lobe of HG modes in most applications like particle manipulation, optical lithography, and the method can be used in other fields like nonlinear frequency conversion and shaping ultrafast optical pulses.

\* Corresponding author.
  *E-mail address*: brijeshsingh@curaj.ac.in (Brijesh Kumar Singh)

*January 18, 2021*

## 1. Introduction

Hermite-Gaussian (HG) optical beam is one of the exact solutions of the free-space paraxial wave equation in Cartesian coordinates systems [1-3]. In the past, much attention has been paid to explore the characteristics of HG laser beams such as excitation in end-pumped solid-state laser [4], terahertz laser [5], incoherent superposition [6], mode transformation [7, 8], nonlinear diffraction [9], the formation of the high-quality beam [10], propagation in the turbulent atmosphere [11-16], mode division multiplexing for free-space optical communications [17-19], atom confinement [20], optical beam trapping [21, 22] and optical lithography [23]. Further, the propagation of the HG beam is studied after modulation by finite energy self- accelerating Airy beam [24] and vortex beam [25]. Recently, a generalized HG beam [26] and non-diffraction property of the truncated HG beam [27] are investigated. High order HG modes are characterized by multiple lobes of different shapes, sizes, and optical energy distributions. The lobes in both x and y directions of HG modes are separated by the particular number of zeros, also known as mode number. Like the Laguerre- Gaussian (LG) beam, the HG beam also shows the shape invariant feature, and its width is a function of the propagation distance. First-order $HG_{10}$ or $HG_{01}$ mode yields two lobes, and $HG_{11}$ yields four lobes of the same size with equal intensity distribution. The symmetry of having all equally intense lobes of the same size breaks down for the modes other than the above HG modes.

In most applications like free-space optical communications and propagation through the turbulent atmosphere of the high order HG beams, the primary focus is on the highest field intensity lobes, which can survive for the longer propagation distance in comparison to that of the low intense lobes of the modes. For example, an elegant HG beam is less affected and less spread by the atmospheric turbulence as well as in free-space optical communications in comparison to the normal HG beam [14]. Further, an elegant HG beam shows enhanced optical manipulation by creating higher trapping forces in comparison to the Gaussian and normal HG beams [28]. In contrast to the high order normal HG beams, the highest intensity lobes concentrate only in the central region while low intense lobes spread toward the outer side in the transverse intensity profiles of the high order elegant HG beams. Further, due to atmospheric turbulence, the high order modes experience less additional broadening beyond diffraction [15]. But the limitation with the high order HG beams is that number of highest intensity lobes is limited to four while numbers of moderate and low intense lobes increase for the orders higher than $HG_{11}$ mode. Further, this uneven intensity distribution among the high order lobes of HG beams restricts them from being utilized simultaneously with equal efficiency. Furthermore, high intense lobes can be used more efficiently than the moderate and low intense lobes of high order HG modes in applications like particle manipulation and optical lithography. Therefore, the question arises whether one can modulate the energy distribution among the lobes of high order HG modes in the selective way that all the lobes become equally high intense lobes and whether low intense lobes can be converted into high or moderate intense lobes and vice-versa.

To address these questions, here we introduce a generic numerical method that modulates the optical energy distributions among the different lobes of high order HG beams in a controlled manner. The optical energy is modulated from high intense lobes to the low intense lobes. In this process, all the lobes become equally intense as well as low intense lobes are converted into the high or moderate intense lobes and vice-versa. Note that wherever we discuss either high or moderate or low intense lobe, that means only the peak intensity value of that lobe is either maximum or in-between or minimum, respectively. In this way, the number of high intense lobes increases in the modulated HG modes, limiting only four in the normal high order HG modes. Hence, all these equally high intense lobes can be used simultaneously with equal efficiency for multiple particle trapping in the optical tweezers. The proposed diffractive optical element (DOE) can be used in optical lithography, nonlinear optics, and modulating ultrafast laser pulses.

## 2. Hermite-gaussian modes

Mathematically, Hermite-Gaussian beam is defined by Hermite polynomial which shows the transverse field distribution with rectangular symmetry and given by following expressions [3],

$$HG_{lm}(x,y,z) = A_{lm} \, G_l\left(\frac{\sqrt{2}x}{W(z)}\right) G_m\left(\frac{\sqrt{2}y}{W(z)}\right) exp\left(-jkz - jk\frac{x^2+y^2}{2R(z)} + j(l+m+1)\zeta(Z)\right), \quad (1)$$

where, $G_l(u) = H_l(u) \exp\left(\frac{-u^2}{2}\right)$ and $G_m(u) = H_m(u) \exp\left(\frac{-u^2}{2}\right)$.

Here, $A_{lm}$ is a constant, $H_l(u)$ and $H_m(u)$ are Hermite polynomials of order $l$ and $m$, respectively with $l, m$ = 0, 1, 2, …, and $\zeta(z)$ is the Gouy phase function. The different HG modes can be numerically simulated directly from Eq. (1). HG beam is known as a self-similar beam, i.e., it retains its shape during propagation. Therefore, its Fourier Transform (FT) also exhibits the transverse energy distribution similar to HG beams given by Eq. (1), and here we utilized this Fourier method for the generation of high order HG beams. The intensity profile at the focal plane ($z = 0$) can be derived via the Fraunhofer diffraction approximation utilizing the two-dimensional (2D) Fourier transform in Cartesian coordinates [29] as given by

$$FT\{HG_{lm}(x,y)\} = \int_{-\infty}^{\infty}\int_{-\infty}^{\infty} HG_{lm}(x,y) \times exp[j2\pi(f_x x + f_y y)\,dxdy] \qquad (2)$$

Where, $f_x$ and $f_y$ are the spatial frequencies along $x$ and $y$ directions, respectively.

## 3. Methods

For controlled modulation of optical energy in the high order HG modes, Eq. (1) is multiplied by an annular binary function $T(r)$ given by [22]

$$T(r) = \begin{cases} -1 & r \leq r_\pi \\ 1 & r_\pi \leq r \leq r_{max} \end{cases} \qquad (3)$$

where, $r_\pi = \gamma r_{max}$, and $r_{max} = d/2$, $d$ is the diameter of the annular mask aperture, and $r$ is the radial coordinate. Here $r_\pi$ is the $\pi$ phase-shifted radial distance, which plays a vital role in the modulation of optical energy, where $\gamma$ is a coefficient which varies as $0 \leq \gamma \leq 1$. The controlled transfer of optical energy among the multiple lobes of high order HG modes is achieved by taking the 2D Fourier Transform [29] of the modulated function $T(r) \times HG_{lm}(x, y)$, i.e., FT $\{T(r) \times HG_{lm}(x, y)\}$. And optical energy among unlike lobes of HG modes is modulated at the Fourier plane and results are numerically simulated by using following parameters at entrance pupil plane: wavelength ($\lambda$) = 532$nm$, the diameter of the mask ($d$) = 3.5$mm$, beam waist of HG beam ($W (z = 0)$) = 1$mm$, Rayleigh range ($Z_R$) = 5.9$m$, pixel size = 8$\mu m$, number of pixels ($Nx, Ny$) = 12288. One dimensional (1D) form of Eq. (3) has been utilized for the generation of the super airy beam [30] and super-oscillatory pulses [31].

## 4. Optical energy modulation in HG modes

## 4.1 HG$_{20}$ mode

The lowest HG modes having uneven peak intensity distributions among the lobes are *HG$_{20}$* and *HG$_{02}$* modes. The phase maps and corresponding Fourier transformed transverse intensity profiles of *HG$_{20}$* mode are shown in the first and second columns of Fig. 1, respectively. From Fig. 1(b), it is seen that all the three lobes are aligned along one row where the outer side two identical lobes are having equal and maximum peak intensities than that of the peak intensity of the central lobe. To confirm the exact peak intensity distributions among each lobe of Fig. 1(b), the 1D intensity distribution is plotted, as shown in Fig. 1(c). The normal phase map of the *HG$_{20}$* mode shown in Fig. 1(a) is modulated by the annular mask given by Eq. (3), and resultant modulated phase maps are shown in Figs. 1 (d) and 1(g) for γ = 0.113 and 0.166, respectively. The transverse intensity profiles are simulated by taking the FT of these modulated phase maps, as shown in Figs. 1(e) and 1(h), respectively. We observed that as we increase the value of γ optical energy is transferred from the outer side lobes to the central lobe, and for γ = 0.113, all the three lobes of *HG$_{20}$* mode have an equal peak intensity at a time as shown in Fig. 1(e). Further, for γ = 0.166 central low intense lobe becomes a high intense lobe while the two identical high intense outer side lobes become together low intense lobes, as shown in Fig. 1(h). This result shows that optical energy distribution in 2D intensity profiles (Fig. 1h) is exactly reversed to that of the normal intensity distribution of *HG$_{20}$* mode shown in Fig. 1(b). All these results are further confirmed from 1D intensity plots of Figs. 1(e) and 1(h), as shown in Figs. 1(f) and 1(i), respectively. Similar results are also observed for *HG$_{02}$* mode for the same γ values as are observed for *HG$_{20}$* mode, respectively. Note that, in all the figures, intensity profiles are normalized from 0 (black) to 1 (white) while phase maps vary from 0 (black) to $\pi$ (white). Scale bars in each normal phase mask of all the figures are also valid for corresponding modulated phase masks of that figure. Scale bar embedded in the normal 2d intensity profile of *HG$_{20}$* mode is authentic for all the 2d intensity profiles of the second column of Figs. 1-7.

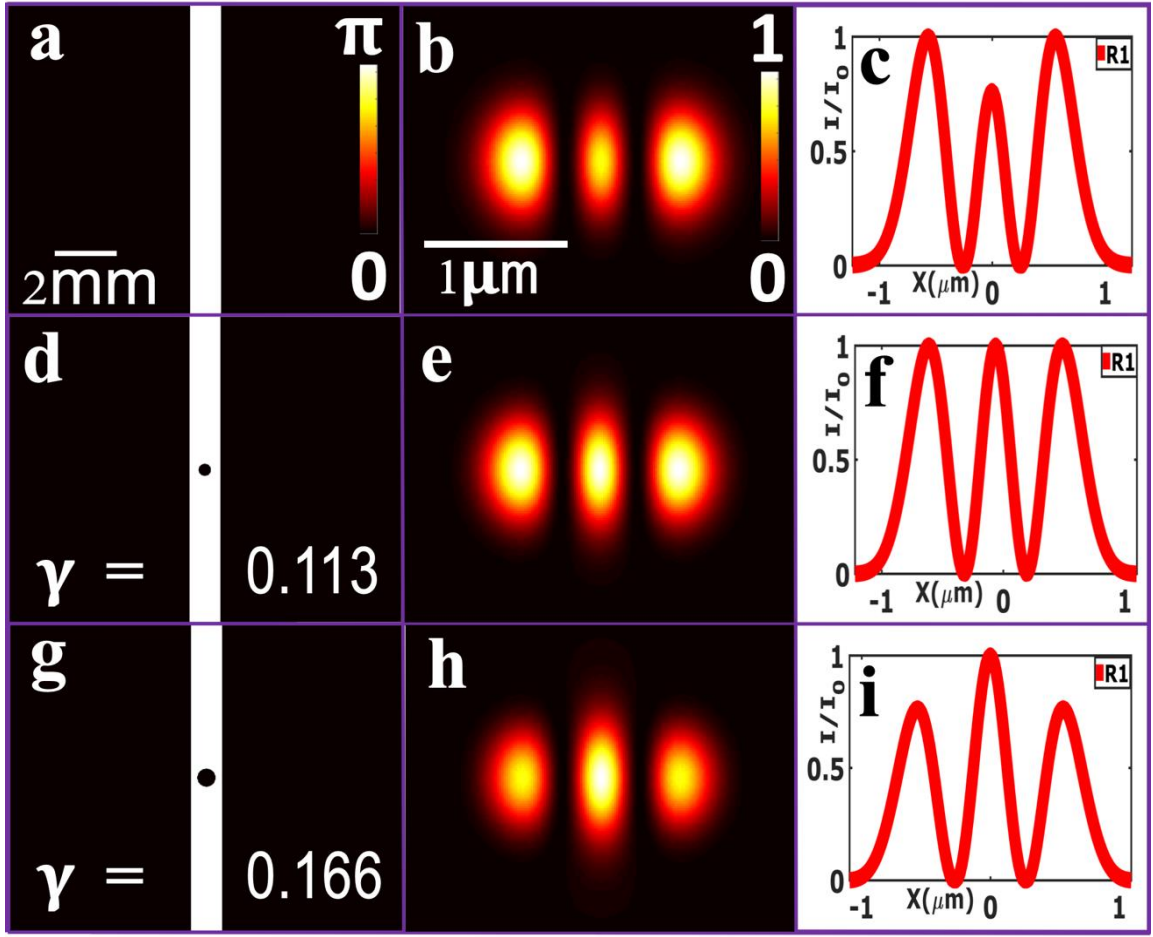

**Fig. 1.** First row is for normal while second, and third rows are for modulated $HG_{20}$ mode for γ = 0.113 and 0.166, respectively. First column (a, d, g) corresponds to the phase maps, second column (b, e, h) corresponds to Fourier transformed 2D intensity profiles of the first column while third column (c, f, i) is 1D plots of the second column, respectively.

### 4.2   $HG_{21}$ mode

Next, the high mode we verified is $HG_{21}$ mode, which phase maps and transverse intensity profiles are shown in the first and second columns of Fig. 2, respectively. From Fig. 2(b), we see that all the six lobes of $HG_{21}$ mode are distributed in the 2 x 3 matrix form, i.e., aligned along two rows and three columns. From the energy distributions, it can be seen that optical energy is not equally distributed among all the six lobes of $HG_{21}$ mode as it consists of four identical high intense corner lobes lie in the first and third columns and two identical low intense central lobes lie in the second column of Fig. 2(b). Here, we found that for γ = 0.293, all the six lobes of $HG_{21}$ mode have equal peak intensity, and for γ = 0.371, both the low intense lobes of the second column become high intense while four high intense corner lobes become low intense lobes as shown in Figs. 2(e) and 2(h), respectively. Result in Fig. 2(h) is opposite to that of the 2D intensity profile of normal $HG_{21}$ mode shown in Fig. 2(b). All these results are further confirmed from their corresponding 1D intensity plots, shown in the third column of Fig. 2, where 1D plots along first (red color) and second (aqua color) rows overlap with each other due to having symmetrical lobes profiles. Similar results are also seen for $HG_{12}$ mode for the same γ values as are observed for $HG_{21}$ mode, respectively.

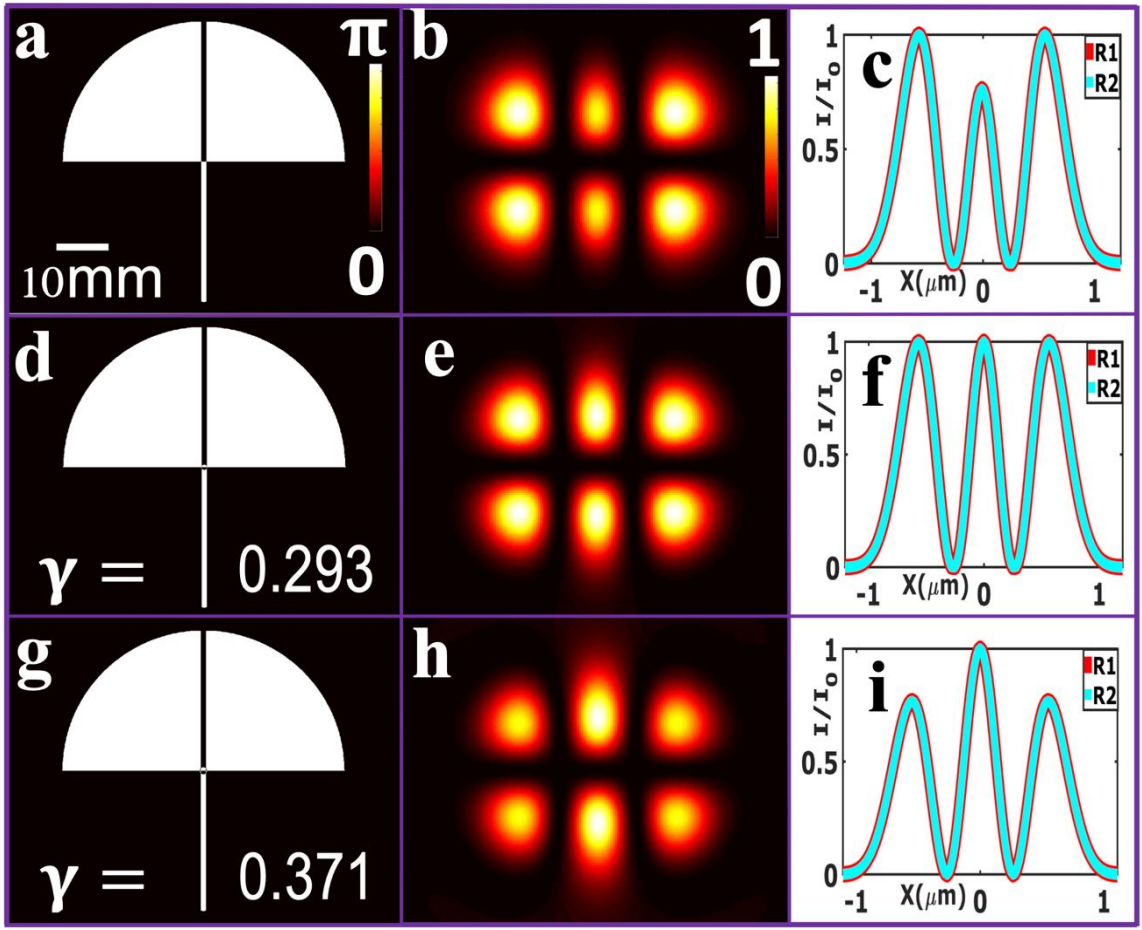

**Fig. 2.** First row is for normal while second, and third rows are for modulated $HG_{21}$ mode for γ = 0.293 and 0.371, respectively. First column (a, d, g) corresponds to the phase maps, second column (b, e, h) corresponds to Fourier transformed 2D intensity profiles of the first column while third column (c, f, i) is 1D plots of the second column, respectively.

## 4.3 HG$_{22}$ mode

Further, we verified the potential of this method for *HG$_{22}$* mode having three different types of intensity distributions among its nine lobes aligned along three rows and three columns of a 3 x 3 matrix, shown in Fig. 3(b). $HG_{22}$ mode consists of four identical high intense corner lobes, four identical moderate intense lobes each lie in-between two high intense lobes, and one low intense lobe at the center of the 2D transverse intensity profile. We observed that for γ = 0.126, all outer eight lobes, i.e., four high intense corner lobes together with four moderate intense lobes of $HG_{22}$ mode, become high intense lobes, as shown in Fig. 3(e). Further, for γ = 0.575, energy is redistributed among the lobes such that all the four moderate intense lobes together with one central low intense lobe become high intense lobes while all four high intense corner lobes together become low intense lobes shown in Fig. 3 (h). For γ = 0.6 modulated intensity profile shown in Fig. 3(k) becomes opposite to that of the normal intensity profile of $HG_{22}$ mode, i.e., low intense central lobe become high intense lobe and all four high intense corner lobes become low intense lobes while peak intensities of all four moderate intense lobes are slightly increased. These results are further confirmed from their 1D intensity plots given in the third column of Fig. 3, where 1D plots along first (red color) and third (indigo color) rows overlap with each other due to having symmetrical lobes profiles.

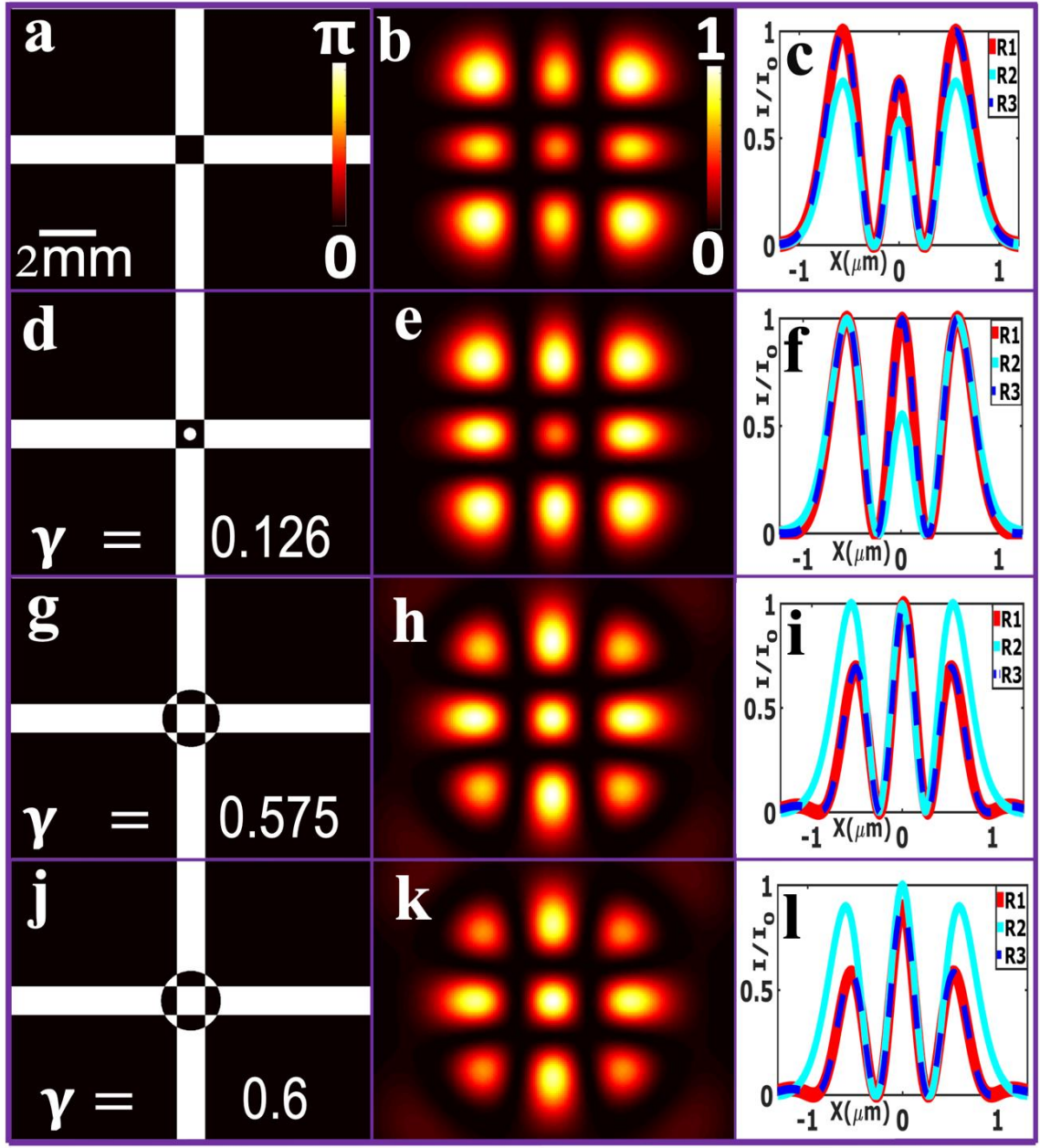

**Fig. 3.** First row is for normal while second to fourth rows are for modulated *HG$_{22}$* mode for γ = 0.126, 0.575, and 0.6, respectively. First column (a, d, g, j) corresponds to the phase maps, second column (b, e, h, k) corresponds to Fourier transformed 2D intensity profiles of the first column while third column (c, f, i, l) is 1D plots of the second column, respectively.

### 4.4 $HG_{03}$ mode

The phase maps and transverse intensity profiles of $HG_{03}$ mode are shown in the first and second columns of Fig. 4, respectively. From Fig. 4(b) it is seen that all the four lobes of the mode are aligned along one column and have two types of intensity distributions. The outer side two identical lobes are highly intense in comparison to that of the two identical central low intense lobes. Here, we found that for γ = 0.253, all the four lobes of $HG_{03}$ mode have equal peak intensity and for γ = 0.313, two low intense central lobes become together high intense lobes while outer side two high intense lobes become together low intense lobes as shown in Figs. 4(e) and 4(h), respectively. Result in Fig. 4(h) is just opposite to that of the 2D intensity profile of normal $HG_{03}$ mode shown in Fig. 4(b). All these results are further confirmed from their corresponding 1D intensity plots, as shown in the third column of Fig. 4. Similar results are also observed for $HG_{30}$ mode for the same γ values as are observed for $HG_{03}$ mode, respectively.

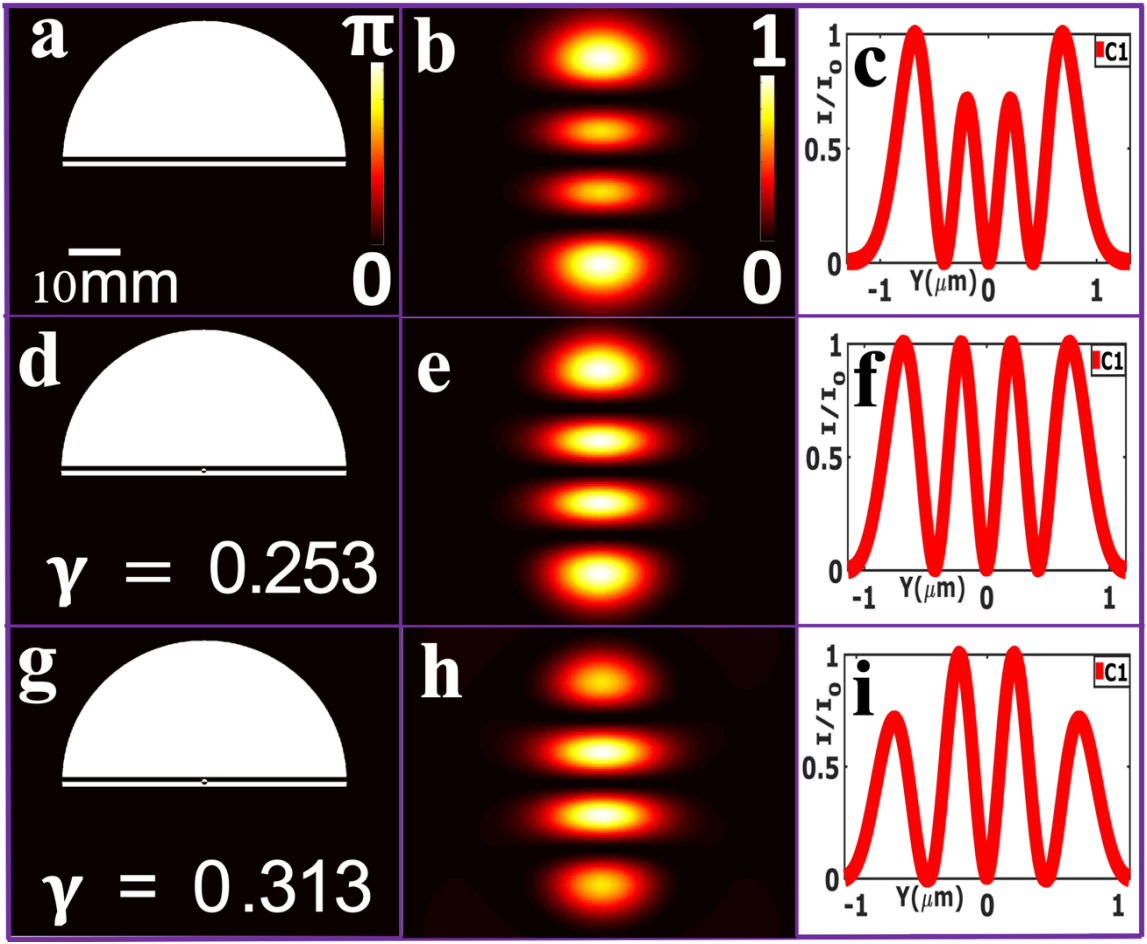

**Fig. 4.** First row is for normal while second, and third rows are for modulated $HG_{03}$ mode for $\gamma$ = 0.253 and 0.313, respectively. First column (a, d, g) corresponds to the phase maps, second column (b, e, h) corresponds to Fourier transformed 2D intensity profiles of the first column while third column (c, f, i) is 1D plots of the second column, respectively.

### 4.5   $HG_{13}$ mode

Next, the high mode we verified is $HG_{13}$ mode, which phase maps and transverse intensity profiles are shown in the first and second columns of Fig. 5, respectively. From Fig. 5(b), we can see that all the eight lobes of $HG_{13}$ mode are distributed in the form of 4 x 2 matrix, i.e., along four rows and two columns. From the energy distributions, it can be seen that $HG_{13}$ mode consists of four identical high intense corner lobes lie in the first and fourth rows, and four identical low intense lobes lie in the second and third rows of Fig. 5(b). We found that for $\gamma$ = 0.4, all the eight lobes of $HG_{13}$ mode have equal peak intensity, and for $\gamma$ = 0.5, four low intense lobes of second and third rows become together high intense lobes while four corner high intense lobes together become low intense lobes, as shown in Figs. 5(e) and 5(h), respectively. Result in Fig. 5(h) is just opposite to that of the 2D intensity profile of normal $HG_{13}$ mode shown in Fig. 5(b). All these results are further confirmed from their corresponding 1D intensity plots, shown in the third column of Fig. 5, where 1D plots along first (red color) and second (aqua color) columns of each 2D intensity profile overlap with each other due to having symmetrical lobes profiles. Similar results are also observed for $HG_{31}$ mode for the same γ values as are observed for $HG_{13}$ mode, respectively.

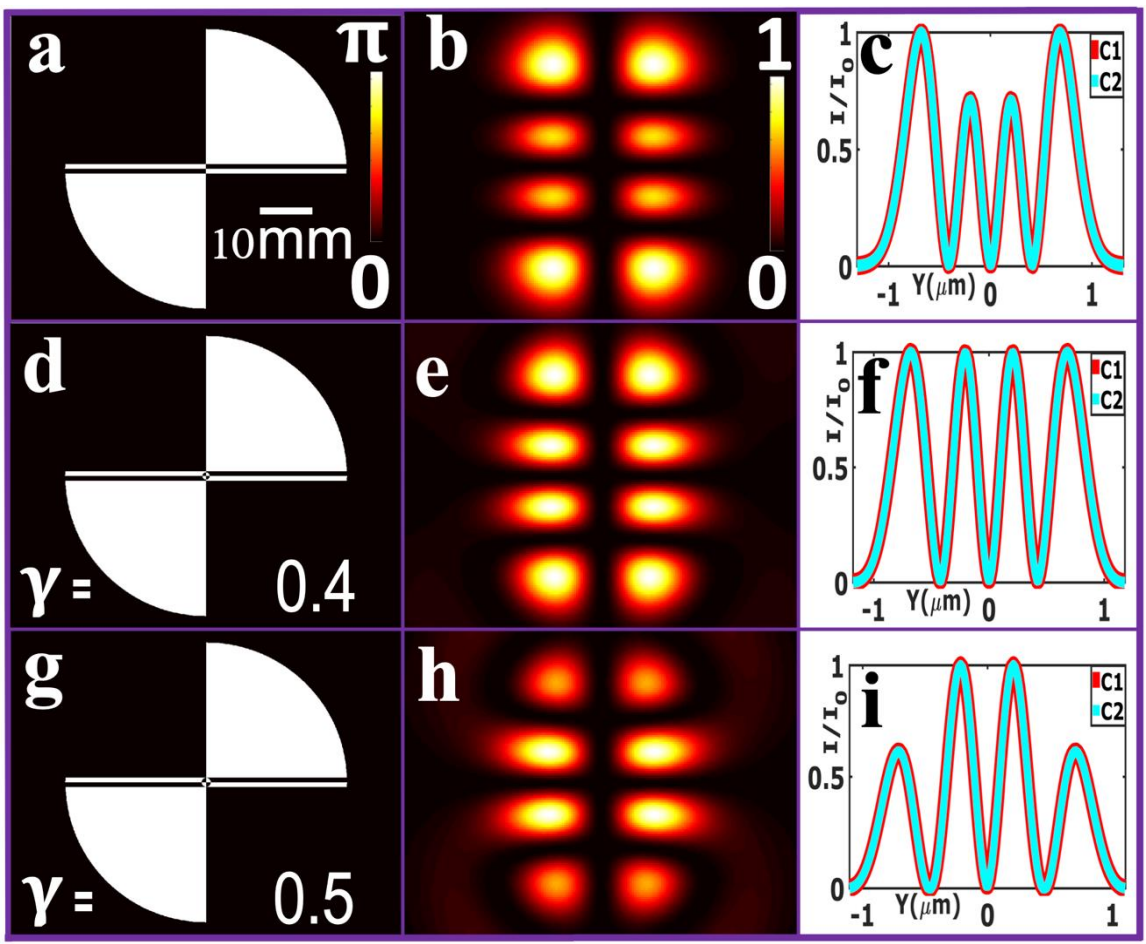

**Fig. 5.** First row is for normal while second, and third rows are for modulated $HG_{13}$ mode for γ = 0.4 and 0.5, respectively. First column (a, d, g) corresponds to the phase maps, second column (b, e, h) corresponds to Fourier transformed 2D intensity profiles of the first column while third column (c, f, i) is 1D plots of the second column, respectively.

## 4.6  $HG_{23}$ mode

Next, we verified $HG_{23}$ mode having complex intensity profiles as it consists of four different types of intensity distribution among its twelve lobes aligned along four rows and three columns of a 4 x 3 matrix shown in Fig. 6(b). Out of twelve lobes, four identical corner lobes lie in the first and third columns are high intense while two identical central lobes lie in the second column are low intense lobes. To make discussion easier, we adopted some terminology for moderate intense lobes. The two identical moderate intense lobes lie in the second column, named as two C2 moderate intense lobes. Similarly, four identical moderate intense lobes out of which two lobes lie in the first column, and the other two lobes lie in the third columns, named as four C13 moderate intense lobes. Two C2 moderate intense lobes are slightly more intense than that of the four C13 moderate intense lobes shown in Fig. 6(b). We observed that for $\gamma$ = 0.23, two C2 moderate intense lobes together with four high intense corner lobes become high intense lobes, i.e., all the six lobes of first and fourth rows, shown in Fig. 6(e). While for $\gamma$ = 0.584, all the six (two C2 and four C13) moderate intense lobes become together high intense lobes and four corner high intense lobes by lowering their peak intensities become moderate intense lobes shown in Fig. 6(h). Further, for $\gamma$ = 0.743, two central low intense lobes together with four C13 moderate intense lobes become high intense lobes, and four high intense corner lobes become low intense lobes shown in Fig. 6(k). Finally, for $\gamma$ = 0.8, the 2D intensity profile is shown in Fig. 6(n) becomes opposite to that of the normal intensity profile of $HG_{23}$ mode shown in Fig. 6(b), i.e., four corner high intense lobes become low intense lobes, and two central low intense lobes become high intense lobes while four C13 moderate intense lobes become more intense in comparison to that of two C2 moderate intense lobes. All these results are further confirmed from their 1D plots given in the third column of Fig. 6, where 1D plots along first (red color) and third (indigo color) columns of each 2D intensity profile overlap with each other due to having symmetrical lobes profiles. Note that all the 1D intensity plots for $HG_{03}$, $HG_{13}$, and $HG_{23}$ modes, lobes lie in first and second columns of each 2D intensity profiles are plotted in red and aqua colors solid lines. In contrast, lobes lie in the third column of each 2D intensity profiles are plotted in indigo color dash lines, respectively. Similar results are also observed for $HG_{32}$ mode for the same $\gamma$ values as are observed for $HG_{23}$ mode, respectively.

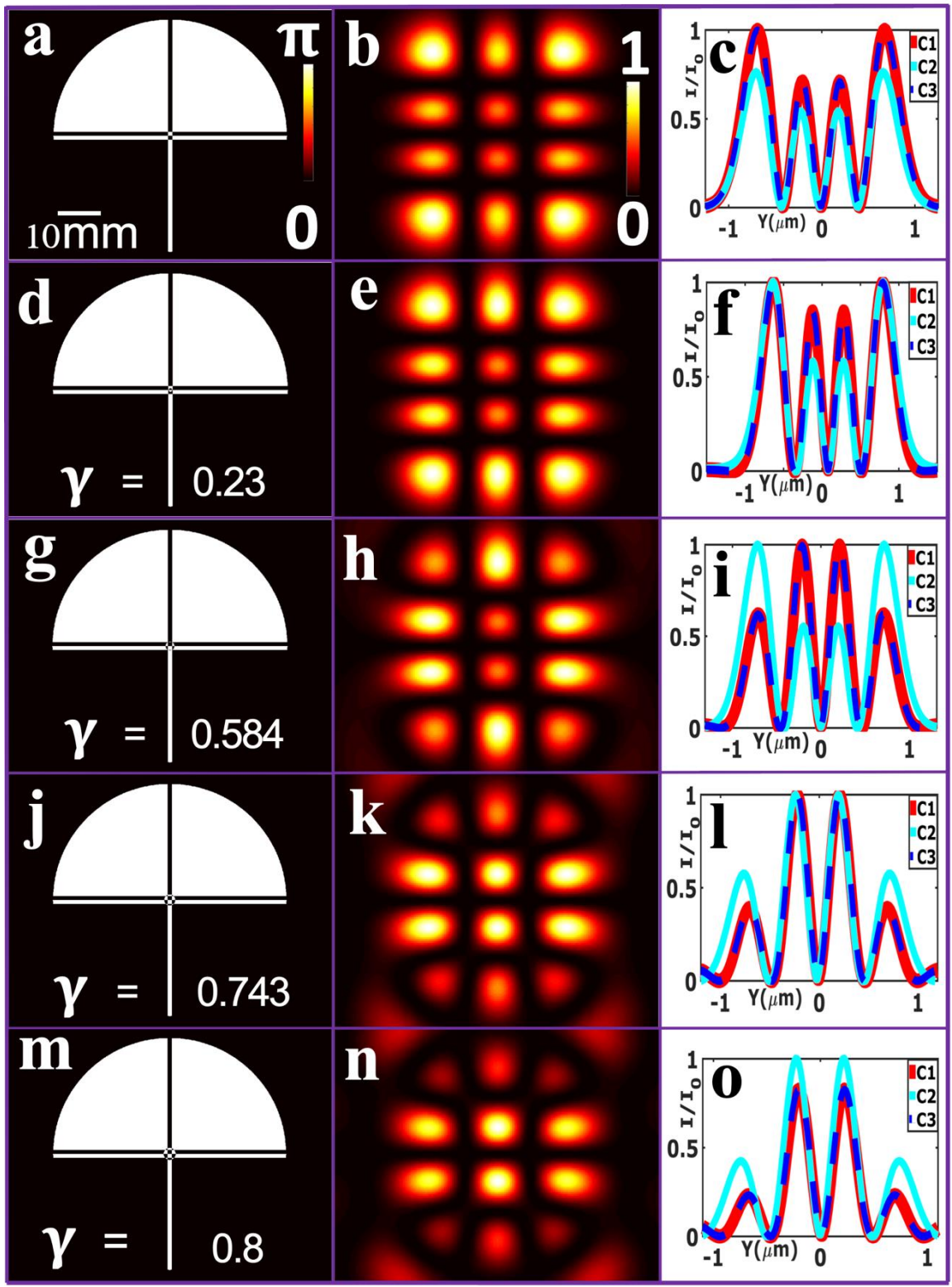

**Fig.6.** First row is for normal while second to fifth rows are for modulated $HG_{23}$ mode for $\gamma$ = 0.23, 0.584, 0.743, and 0.8, respectively. First column (a, d, g, j, m) corresponds to the phase maps,

second column (b, e, h, k, n) corresponds to Fourier transformed 2D intensity profiles of the first column while third column (c, f, i, l, o) is 1D plots of the second column, respectively.

### 4.7  $HG_{33}$  mode

Finally, the potential of this method is tested to the high order mode of HG beam having a large number of lobes, i.e., $HG_{33}$ mode. It consists of three different types of intensity distribution among its sixteen lobes aligned along four rows and four columns of a 4 x 4 matrix shown in Fig. 7(b). Four identical corner lobes are high intense while four identical central lobes lie along two diagonals of the matrix are low intense lobes. Further, eight identical moderate intense lobes lie in between two corner high intense lobes. We observed that for γ = 0.346, all the outer twelve lobes, i.e., four high intense corner lobes together with eight moderate intense lobes, become high intense lobes shown in Fig. 7(e), while for γ = 0.581, all the eight moderate intense lobes become high intense lobes and four corner high intense lobes by lowering their peak intensities become nearly equal to that of four central low intense lobes shown in Fig. 7(h). As we further increase the value to γ = 0.75, four central low intense lobes gained optical energy to become moderate intense lobes, and four high intense corner lobes become low intense lobes whereas all the eight moderate intense lobes remain high intense lobes shown in Fig. 7 (k). In the same direction, for γ = 0.808, four identical central low intense lobes together with eight moderate intense lobes become high intense lobes, while four high intense corner lobes remain low intense lobes shown in Fig. 7(n). Furthermore, a modulated intensity profile for γ = 0.88 shown in Fig. 7(q) becomes opposite to that of the normal intensity profile of $HG_{33}$ mode shown in Fig. 7(b), i.e., four corner high intense lobes become low intense lobes, and four central low intense lobes become high intense lobes while eight moderate intense lobes remain moderate intense lobes. All these results are further confirmed from their 1D plots given in the third column of Fig. 7. For γ ≥ 0.6 additional low-intensity noises appear at the outer side of four high intense corner lobes. Note that, in all the 1D intensity plots for all the HG modes (except $HG_{03}$, $HG_{13}$, and $HG_{23}$ modes), lobes lie in first and second rows of each 2D intensity profiles are plotted in red and aqua colors solid lines, while lobes lie in third and fourth rows of each 2D intensity profiles are plotted in indigo and yellow colors dash lines, respectively.

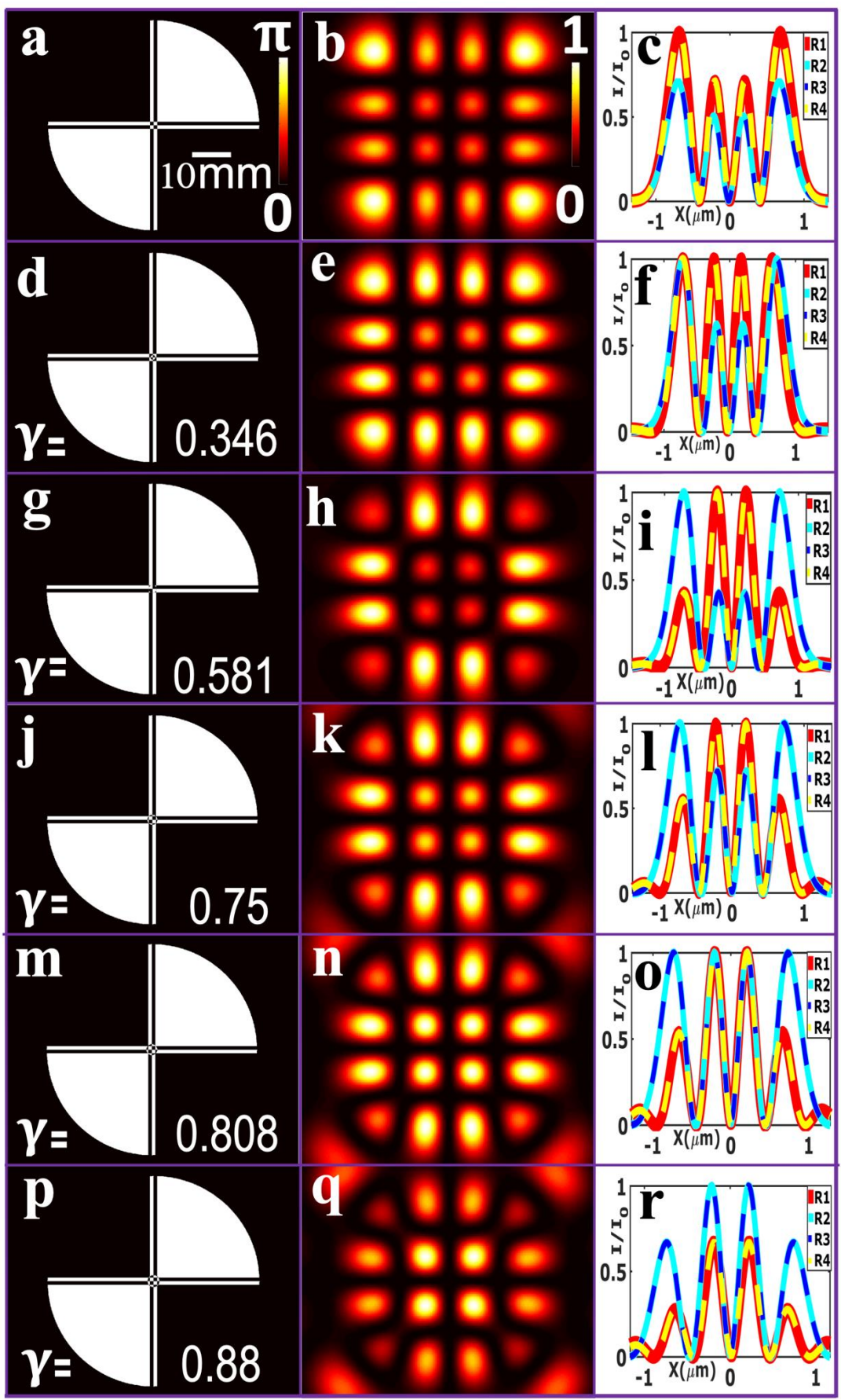

**Fig. 7.** First row is for normal while second to sixth rows are for modulated $HG_{33}$ mode for γ = 0.346, 0.581, 0.75, 0.808, and 0.88, respectively. First column (a, d, g, j, m, p) corresponds to the phase maps, second column (b, e, h, k, n, q) corresponds to Fourier transformed 2D intensity profiles of the first column while third column (c, f, i, l, o, r) is 1D plots of the second column, respectively.

## 5. Discussion

As we can see from Eq. (3) that annular phase mask consists of two binary zones where the inner zone is modulated by an extra π phase shift in comparison to the outer zone of the mask. The phase difference between inner and outer zones light waves produces destructive interference at the focal point. While at its vicinity, waves are in phase, and thus a constructive field distribution appears. Constructive and destructive interference at the focal plane modulates the energy of the HG modes. Further, γ is an important parameter, and its value in the phase mask plays a vital role in the redistribution of optical energy in the diffraction pattern. We see that the values of γ are different for equally intense lobes for different high order HG modes. To understand this manifestation, we need to look at the functions used at the entrance pupil plane where different high order HG modes $HG_{lm}$ (x, y) are modulated by the annular phase mask $T(r)$, i.e., given by $T(r) \times HG_{lm}(x, y)$. Redistribution of optical energy among the multiple lobes of high order HG modes is achieved by taking the 2D Fourier Transform [29] of the modulated function $T(r) \times HG_{lm}$ (x, y), i.e., mathematically given by FT{$T(r) \times HG_{lm}$ (x, y)} = FT{$T(r)$} ⊗ FT {$HG_{lm}$ (x, y)}, where "⊗" symbol is a convolution operator. The analytical expression of the complex field $\tilde{E}(r_\pi, \rho, z)$ of the modulated HG beams at the Fourier transform plane is given by

$$\tilde{E}(r_\pi, \rho, z) = \left[ \frac{r_{max}}{\rho} J_1(2\pi r_{max} \rho) - 2 \frac{r_\pi}{\rho} J_1(2\pi r_\pi \rho) \right] \otimes HG_{l,m}(\rho, z) \qquad (4).$$

Where $J_1$ is first-order Bessel function of the first kind and $\rho$ is the radial coordinate at the Fourier plane. It shows that high order HG modes convolved at the Fourier plane with the mask's diffraction pattern, i.e. the superposition of two Bessel functions. Hence, depending on the shape, size, and intensity distribution of the different high order HG modes, the value of γ is also different to make all the lobes equally intense or for other configurations (i.e., low, and moderate intense lobes) of different HG modes. Further, the annular phase mask $T(r)$, when illuminated with a uniform input plane wave, is known to produce a super-resolution focal spot at the central region with the intense side lobes [22]. For high values of γ, these side lobes become more prominent in comparison to that of the central focal spot. When the phase mask is illuminated with the desired function as in our case high order HG modes at the entrance pupil plane, because of the different shapes and sizes of high order HG modes, side lobes are differently modulated by the various high order HG modes at the Fourier plane. Therefore, unwanted low intense side lobes at the corners of the 2d diffraction patterns are seen, which are less or more prominent for the same value of γ for different high order HG modes.

Furthermore, we see that the sensitivity of the focal plane after diffraction depends on the focal length of the Fourier transforming lens. If we use a tight focusing lens (high numerical aperture (NA)), then the diffraction pattern is susceptible to the focal plane. At the same time, for loose focusing lens (low NA) same diffraction pattern can be observed slightly away from the focal plane. The effective numerical aperture in our case is NA $_{(eff)}$ = 0.6 < 0.7; therefore, the impact of input polarization of the beam on the focal plane intensity profile can be neglected. Consequently, we used the scalar diffraction theory to simulate the intensity profile in the focal plane and have considered linear polarization only. The polarization effect would be more prominent if we use tight focusing. Further, we notice that for high values of γ, peak intensity locations of different lobes lie in specific rows/columns in the modulated 2D intensity profiles are not aligned precisely either in a single row or single column. Therefore, to resolve this issue, we retrieved the data corresponding to the peak intensity' row/column of each lobe aligned in a specific row/column of 2d intensity profiles and transferred into a new matrix in the form of a single row/column, respectively. This new matrix enables all the peak intensity locations of each lobe to be aligned in a single row/column. In this way, in Figs. 1-3, and Fig. 7, 1D plots are plotted only along the peak intensity values of each lobe in a particular row while in Figs. 4-6, 1D plots are plotted only along the peak intensity values of each lobe lies in a specific column. The Fourier transform of π phase-modulated HG masks results in appropriate diffraction patterns at the Fourier plane. That redistribute optical energy among multiple lobes of high order HG modes in a particular way such that moderate and low intense lobes are converted to high intense lobes and vice-versa, for specific values of γ.

## 6. Conclusions

In conclusion, we numerically presented here a systemic method that modulated optical energy among the multiple lobes of high order HG modes in a controlled manner. We observed that for a specific value of γ all the lobes converted to high intense lobes for the high order modes having only two types (high and low intense lobes) of the intensity distribution. Further, modulated intensity profiles become opposite to their normal intensity profiles of HG modes by converting all low intense lobes together to high intense lobes and vice-versa. Furthermore, as we increase the order of HG modes, the number of lobes in each mode increases, which results in more complex energy distribution among the lobes as the different types of moderate intense lobes increase in 2D transverse intensity profiles. This method increases the number of high intense lobes in the modulated HG beams, limited to only four in the normal high order HG modes. As we applied this numerical method to these high order HG modes, we observed that sets of lobes having different peak intensities become together either high intense or low intense or moderate intense lobes, for particular values of γ in a controlled manner. Note that all the equally intense lobes transformed altogether to other intensity values. We envisage that such controlled modulation of optical energy among the lobes may be used for selective optical trapping. Further, all equally high intense lobes can be utilized simultaneously for multiple particle trapping with equal efficiency in optical tweezers as well as in optical lithography. Finally, the method of modulating optical energy among the lobes shown here can be used in other fields, for example, nonlinear frequency conversion, plasmonics as well as in the time domain for modulating light pulses.

**Acknowledgment**

Authors would like to acknowledge University Grant Commission (UGC) of India and Science and Engineering Research Board (CRG/2019/001187) for their financial support.

**Disclosures.** The authors declare no conflicts of interest.

**References**

[1]     H. Kogelnik, T. Li, Laser Beams and Resonators, Applied optics. 5(1966) 1550-1567. https://doi.org/10.1364/AO.5.001550

[2]     A. E. Siegman, Lasers, University Science Books, Mill Valley, California, 1968.

[3]     B.E.A. Saleh, M.Carl. Teich, Fundamentals of photonics, Wiley, 1991.

[4]     H. Laabs, B. Ozygus, Excitation of Hermite Gaussian modes in end-pumped solid-state lasers via off-axis pumping, Optics & laser technology.28(1996) 213-214. https://doi.org/10.1016/0030-3992(95)00087-9

[5]     E. Abramochkin, V. Volostnikov, Beam transformations and nontransformed beams, Optics Communications. 83 (1991) 123–135. https://doi.org/10.1016/0030-4018(91)90534-K .

[6]     F. Gori, M. Santarsiero, R. Borghi, G. Guattari, Intensity-based modal analysis of partially coherent beams with Hermite-Gaussian modes, Optics letter. 23(1998) 989-991. https://doi.org/10.1364/OL.23.000989

[7]     M.W. Beijersbergen, L. Allen, H.E.L.O. van der Veen, J.P. Woerdman, Astigmatic laser mode converters and transfer of orbital angular momentum, Optics Communications. 96 (1993) 123-132. https://doi.org/10.1016/0030-4018(93)90535-D

[8]     A.N. Agafonov, Yu.Yu. Choporova, A.K. Kaveev, B.A. Knyazev, G.I. Kropotov, V.S. Pavelyev, K.N. Tukmakov, B.O. Volodkin, Control of transverse mode spectrum of Novosibirsk free electron laser radiation, Applied Optics. 54 (2015) 3635–3639. https://doi.org/10.1364/ao.54.003635.

[9]     K. Kalinowski, A. Shapira, A. Libster-Hershko, A. Arie, Nonlinear diffraction from high-order Hermite–Gauss beams, Optics Letters. 40 (2015) 13-16. https://doi.org/10.1364/ol.40.000013
.

[10]    G. Machavariani, N. Davidson, A.A. Ishaaya, A.A. Friesem, E. Hasman, Efficient formation of a high-quality beam from a pure high-order Hermite-Gaussian mode, Optics letters 27(2002) 1501-1503. https://doi.org/10.1364/OL.27.001501

[11]    Y. Baykal, Correlation and structure functions of Hermite-sinusoidal-Gaussian laser beams in a turbulent atmosphere, Journal of the Optical Society of America A 21(2004) 1290-1299. https://doi.org/10.1364/JOSAA.21.001290

[12]    H. Tanyer E. ˇlu, Hermite-cosine-Gaussian laser beam and its propagation characteristics in turbulent atmosphere Journal of the Optical Society of America A (2005) 1527-1535. https://doi.org/10.1364/JOSAA.22.001527

[13]    X. Ji, X. Chen, B. Lü, Spreading and directionality of partially coherent Hermite-Gaussian beams propagating through atmospheric turbulence, Journal of the Optical Society of America A (2007) 21-28. https://doi.org/10.1364/JOSAA.25.000021

[14]    Yuan Yangsheng, Cai Yangjian, Qu Jun, Eyyuboğlu Halil T, Baykal Yahya, Average intensity and spreading of an elegant Hermite–Gaussian beam in turbulent atmosphere, Optics Express (2009). 139-1130 https://doi.org/10.1364/OE.17.011130

[15]    Y.V. Gilchrest, Turbulence induced beam spreading of higher order mode optical

waves, Optical Engineering. 41 (2002) 1097-1103.https://doi.org/10.1117/1.1465427.

[16]   M.A. Cox, L. Maqondo, R. Kara, G. Milione, L. Cheng, A. Forbes, The Resilience of Hermite-and Laguerre-Gaussian Modes in Turbulence, Journal of Lightwave Technology. 37 (2019) 3911–3917. https://doi.org/10.1109/JLT.2019.2905630.

[17]   B. Ndagano, N. Mphuthi, G. Milione, A. Forbes, Comparing mode-crosstalk and mode-dependent loss of laterally displaced orbital angular momentum and Hermite–Gaussian modes for free-space optical communication, Optics Letters. 42 (2017) 4175-4178. https://doi.org/10.1364/ol.42.004175.

[18]   K. Pang, H. Song, Z. Zhao, R. Zhang, H. Song, G. Xie, L. Li, C. Liu, J. Du, A.F. Molisch, M. Tur, A.E. Willner, 400-Gbit/s QPSK free-space optical communication link based on four-fold multiplexing of Hermite–Gaussian or Laguerre–Gaussian modes by varying both modal indices, Optics Letters. 43 (2018) 3889. https://doi.org/10.1364/ol.43.003889.

[19]   A. Amphawan, S. Chaudhary, T.K. Neo, Hermite-gaussian mode division multiplexing for free-space optical interconnects, Advanced Science Letters. 21 (2015) 3050–3053. https://doi.org/10.1166/asl.2015.6532.

[20]   T.P. Meyrath, F. Schreck, J.L. Hanssen, C. -S Chuu, M.G. Raizen, A high frequency optical trap for atoms using Hermite-Gaussian beams, Optics Express (2005) 2843-2851. https://doi.org/10.1364/OPEX.13.002843

[21]   A.P. Porfirev, R. v. Skidanov, Optical trapping and manipulation of light-absorbing particles by means of a Hermite–Gaussian laser beam, Journal of Optical Technology. 82 (2015) 587–591. https://doi.org/10.1364/jot.82.000587.

[22]   B.K. Singh, H. Nagar, Y. Roichman, A. Arie, Particle manipulation beyond the diffraction limit using structured super-oscillating light beams, Light: Science and Applications. 6 (2017) 1-7. https://doi.org/10.1038/lsa.2017.50.

[23]   S. Lightman, R. Gvishi, G. Hurvitz, A. Arie, Comparative analysis of direct laser writing and nanoimprint lithography for fabrication of optical phase elements, Applied Optics. 55 (2016) 9724-9730. https://doi.org/10.1364/AO.55.009724.

[24]   J. Yu, S. Xiao, L. Yao, S. Liu, J. Li, Propagation of the finite energy Airy-Hermite-Gaussian beams in uniaxial crystals orthogonal to the optical axis, Journal of Modern Optics. 64 (2017) 616–623. https://doi.org/10.1080/09500340.2016.1254829.

[25]   V.V. Kotlyar, A.A. Kovalev, Hermite–Gaussian modal laser beams with orbital angular momentum, Journal of the Optical Society of America A. 31 (2014) 274-282. https://doi.org/10.1364/josaa.31.000274.

[26]   Y. Wang, Y. Chen, Y. Zhang, H. Chen, S. Yu, Generalised Hermite-Gaussian beams and mode transformations, Journal of Optics (United Kingdom). 18 (2016) 1-6. https://doi.org/10.1088/2040-8978/18/5/055001.

[27]     A. Bencheikh, A. Forbes, The non-diffracting nature of truncated Hermite–Gaussian beams, Journal of the Optical Society of America A. 37 (2020) C1-C6. https://doi.org/10.1364/josaa.385913.

[28]     C. Alpmann, C. Schöler, C. Denz, Elegant Gaussian beams for enhanced optical manipulation, Applied Physics Letters. 106 (2015) 241102(1-5). https://doi.org/10.1063/1.4922743.

[29]     Joseph W. Goodman, Introduction to Fourier Optics, McGraw-Hill: San Francisco, CA, USA 1996.

[30]     B.K. Singh, R. Remez, Y. Tsur, A. Arie, Super-Airy beam: self-accelerating beam with intensified main lobe, Optics Letters. 40 (2015) 4703-4706. https://doi.org/10.1364/ol.40.004703.

[31]     Y. Eliezer, B.K. Singh, L. Hareli, A. Bahabad, A. Arie, Experimental realization of structured super-oscillatory pulses, Optics Express. 26 (2018) 4933-4941. https://doi.org/10.1364/oe.26.004933.